\documentclass{article}  

\usepackage{graphicx}

\usepackage{geometry}
\usepackage{txfonts}
\usepackage{amssymb}
\usepackage{amstext}
\usepackage{natbib}

\usepackage{hyperref}
\newcommand{\fig}[1]{Fig.~\ref{#1}}

\begin{document} 


\Large
\begin{center}\textbf{Evidence for auroral influence on Jupiter's nitrogen and oxygen chemistry revealed by ALMA}\end{center}\normalsize

\large\noindent T. Cavali\'e$^{1,2}$, L. Rezac$^{3}$, R. Moreno$^{2}$, E. Lellouch$^{2}$, T. Fouchet$^{2}$, B. Benmahi$^{1}$, T. K. Greathouse$^{4}$, J. A. Sinclair$^{5}$, V. Hue$^{4}$, P. Hartogh$^{3}$ M. Dobrijevic$^{1}$, N. Carrasco$^{6}$, Z. Perrin$^{6}$\normalsize\\
\vspace{0.2cm}

\noindent$^1$Laboratoire d'Astrophysique de Bordeaux, Univ. Bordeaux, CNRS, B18N, all\'ee Geoffroy Saint-Hilaire, 33615 Pessac, France (ORCID: 0000-0002-0649-1192, email: thibault.cavalie@u-bordeaux.fr)\\ 
$^2$LESIA, Observatoire de Paris, PSL Research University, CNRS, Sorbonne Universit\'es, UPMC Univ. Paris 06, Univ. Paris Diderot, Sorbonne Paris Cit\'e, Meudon, France\\
$^3$ Max-Planck-Institut f\"ur Sonnensystemforschung, 37077 G\"ottingen, Germany\\
$^4$ Southwest Research Institute, San Antonio, TX 78228, United States\\
$^5$ Jet Propulsion Laboratory, California Institute of Technology, 4800 Oak Grove Drive, Pasadena, CA 91109, USA\\
$^6$ LATMOS, CNRS, UVSQ Universit\'e Paris-Saclay, Sorbonne Universit\'e ; 11 boulevard d'Alembert, 78280 Guyancourt, France\\

\vspace{0.2cm}
\noindent\textbf{Received:} 14 October 2022\\
\noindent\textbf{Accepted:} 26 May 2023\\
\noindent\textbf{Published:} 6 July 2023\\
\vspace{0.2cm}

\noindent\textbf{DOI:} https://doi.org/10.1038/s41550-023-02016-7\\
\vspace{0.5cm}

\section*{Abstract}
The localized delivery of new long-lived species to Jupiter's stratosphere by comet Shoemaker-Levy 9 in 1994 opened a window to constrain jovian chemistry and dynamics by monitoring the evolution of their vertical and horizontal distributions. However, the spatial distributions of CO and HCN, two of these long-lived species, had never been jointly observed at high latitudinal resolution. ALMA observations of HCN and CO in March 2017 show that CO was meridionally uniform and restricted to pressures lower than 3$\pm$1 mbar. HCN shared a similar vertical distribution in the low-to-mid latitudes, but was surprisingly depleted at pressures between 2$_{-1}^{+2}$  and 0.04$_{-0.03}^{+0.07}$ mbar in the aurora and surrounding regions, resulting in a drop by two orders of magnitude in column density. We propose that heterogeneous chemistry bonds HCN on large aurora-produced aerosols at these pressures in the jovian auroral regions causing the observed depletion.

\section{Introduction}
Jupiter is an archetype for gas giants (in the solar system and in extrasolar systems). Like the other giants in the Solar System, it is a fast rotator (9.9-hr period). The sheer size and angular speed induce a very complex atmospheric structure, circulation and meteorological system such that the dynamics and chemistry at work in Jupiter's atmosphere are still poorly understood. 

Mainly composed of hydrogen and helium, giant planets also contain 0.2-4\% methane. Its photolysis at high altitude initiates the production of more complex hydrocarbons (1). Giant planets also capture external material, in the form of infalling comets, interplanetary dust, or gas and grains from their ring and satellite systems (2), contributing to enhance the complexity of the chemistry of their atmospheres. While 1D photochemical models generally succeed in explaining the disk-averaged abundances of these species (1), no model has yet managed to reproduce the meridional distributions and the temporal evolution of their hydrocarbons and other trace species (e.g., 3).

A rare event at Jupiter can help us better constrain both chemistry and dynamics in Jupiter's atmosphere. July 1994 saw the first extraterrestrial collision in the Solar System witnessed from Earth with multi-spectral observations. The 21 fragments of comet Shoemaker Levy 9 (SL9) spectacularly impacted Jupiter in its southern hemisphere near 44$^\circ$S (4), increased its temperature locally (5, 6) and left the planet with visible dark scars for weeks (e.g., 7, 8). Even more significant on the long term, the SL9 impacts produced a series of species previously undetected in the stratosphere, like CO, HCN, CS, H$_2$O, and CO$_2$ (9-11), probably from shock chemistry during the impacts (12), except CO$_2$ which probably formed subsequently from atmospheric photochemistry (13). These species were deposited at $\sim$0.1 mbar during the splashback of the impact plumes (14, 15), and subsequently spread in Jupiter's stratosphere. While they contaminated all longitudes at the impact site latitude within approximately one year (16), the meridional and vertical diffusion occurred on longer timescales. (15) predicted that meridional eddy diffusion would require more than a decade to see abundances uniformly mixed. This was confirmed by observations of HCN and CO$_2$ in 1995 and 2000 (17, 18) and the long-term monitoring of the H$_2$O vertical distribution (19, 20). Given that CO, HCN, CS and H$_2$O have sufficiently long chemical lifetimes ($>$10 years), their deposition by SL9 in 1994 offers us a unique opportunity to study the temporal evolution of their distributions over several years and now decades. This is a powerful tool to better understand the chemistry and dynamics of Jupiter's stratosphere.

CO, HCN, and CS have been monitored ever since the SL9 impacts. (15) have mapped their distributions until 1998 with a moderate spatial resolution of 1/3--1/4 of the planet diameter. Further monitoring over the following decade consistently showed a slow decrease of their disk-averaged masses, with decay factors ranging from 5 to 15 between 1998 and 2006 for the different species. Despite the slow decrease of their abundances, CO, HCN, and CS, are stable enough that they can be used as dynamical tracers in the Jovian atmosphere, using their vertical and latitudinal distributions. Based on observations over 1994-1997 with the IRAM-30m and the IRTF, (15) and (17) obtained the first estimates on latitudinal eddy mixing. (18) revealed that, in addition to a large S-N hemispheric asymmetry caused by the slow diffusion from the impact sites toward northern latitudes, HCN showed an abrupt decrease southward of 45$^\circ$S and northward of 50$^\circ$N, that was interpreted as a ``dynamical barrier'' isolating the high latitudes from other latitudes.

Most remarkably, CO$_2$, observed simultaneously with HCN by Cassini, revealed a strikingly different distribution from HCN, peaking at the South Pole instead (18). The difference is difficult to understand as CO$_2$ is thought to be a daughter molecule of CO (from CO $+$ OH $\rightarrow$ CO$_2$ $+$ H, where OH is produced from H$_2$O) (13), which was produced by the SL9 impacts, similar to HCN. (18) explored models in which the CO$_2$ polar excess was associated with the conversion of precipitating oxygen-bearing material to CO$_2$, but did not find this to be a promising scenario. (18) also investigated various horizontal transport models combining latitudinal advection and strongly latitude-dependent eddy mixing and tentatively concluded that the HCN and CO$_2$ were affected by meridional transport in opposite directions (equatorward and poleward, respectively), implying that the two species resided at different atmospheric levels. The conclusions were however strongly hampered by the lack of information on the behavior of CO.

\section{Observations}
\begin{figure*}[!h]
\begin{center}
   \includegraphics[width=15cm,keepaspectratio]{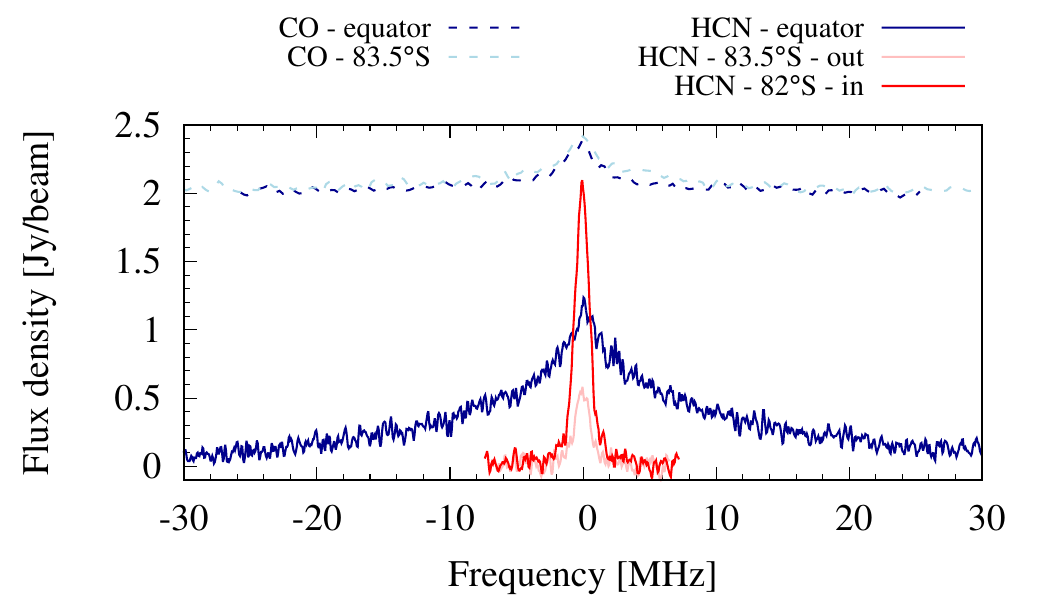}
\end{center}
\caption{Sample of HCN and CO spectra observed with ALMA in Jupiter on March 22nd, 2017. Their corresponding latitude on Jupiter is given (``in'' stands for inside the aurora and is on the 350W limb, and ``out'' stands for outside the aurora and is on the 170W limb). All spectra have been centered around their central frequency for an easier comparison and the CO spectra have been shifted by 2 Jy/beam.}
\label{Fig1} 
\end{figure*}

We used ALMA and Gemini/TEXES observations (21-23) to retrieve the spatial distribution of HCN and CO in Jupiter's stratosphere to better understand their temporal evolution. The ALMA observations consist of high signal-to-noise ratio maps of the HCN (4-3) and CO (3-2) spectral emissions and were recorded on March 22nd, 2017. A sample of spectra is shown in \fig{Fig1}. The latitudinal resolution was $\sim$2$^\circ$ at low latitudes and the spectral resolving power 3$\times$10$^6$. The Gemini/TEXES observations of the CH$_4$ band at 7.8 $\mu$m were recorded on March 14, 16 and 20, 2017 by (22) for the low-to-mid latitudes and on March 17--19, 2017 by (23) for the high latitudes. These were used to retrieve the temperature field between 1 $\mu$bar and 50 mbar with uncertainties $<$ 2 K. The northern polar region was mapped on March 17 and 19, while the southern polar region was only mapped on March 18. We filled latitude and longitude coverage gaps by interpolating between the data points (see \fig{ExtDataFig1}). We used the combination of temperature maps obtained nearly simultaneously to the ALMA observations, a forward radiative transfer modeling and a retrieval algorithm (see Methods) to retrieve the vertical profiles of HCN and CO as a function of latitude from spectra observed at the limb of Jupiter (see \fig{ExtDataFig2}). These profiles are parametrized with three independent numbers, which essentially mirror the fact that the line amplitude depends on the mole fraction in the upper stratosphere and that the whole lineshape controls the pressure above which the mole fraction drops and the slope of the decrease. The vertical profile retrievals have a very limited sensitivity to the temperature uncertainties ($<$ 1\%), because the lines are not optically thick (i.e., $\tau$$=$ 0.2 for CO, 0.4 for HCN at high latitudes, and 1.6 for HCN in the low-to-mid latitudes).

\section{The spatial distribution of CO and HCN}
\begin{figure*}[!h]
\begin{center}
   \includegraphics[width=15cm,keepaspectratio]{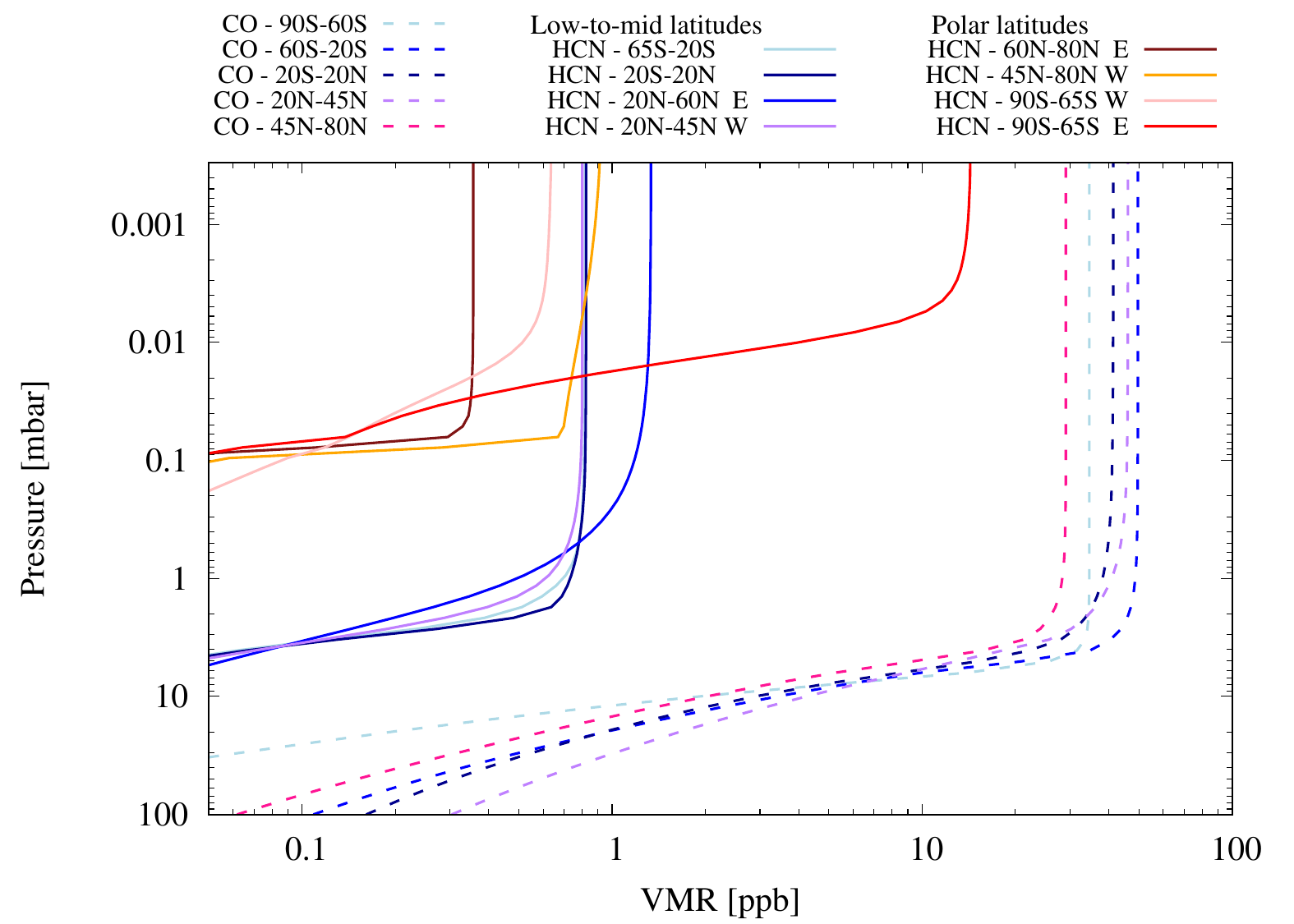}
\end{center}
\caption{HCN and CO volume mixing ratio (VMR) vertical profiles in Jupiter's stratosphere. The HCN (solid lines) and CO (dashed lines) retrieved profiles have been averaged by latitude bins for clarity. For HCN, we differentiate profiles seen on each limb for some particular latitude bins: E stands for the eastern limb (350W longitude) and W for the western one (170W longitude). The vertical range corresponds to the range of vertical sensitivity.}
\label{Fig2} 
\end{figure*}

\begin{figure*}[!h]
\begin{center}
   \includegraphics[width=15cm,keepaspectratio]{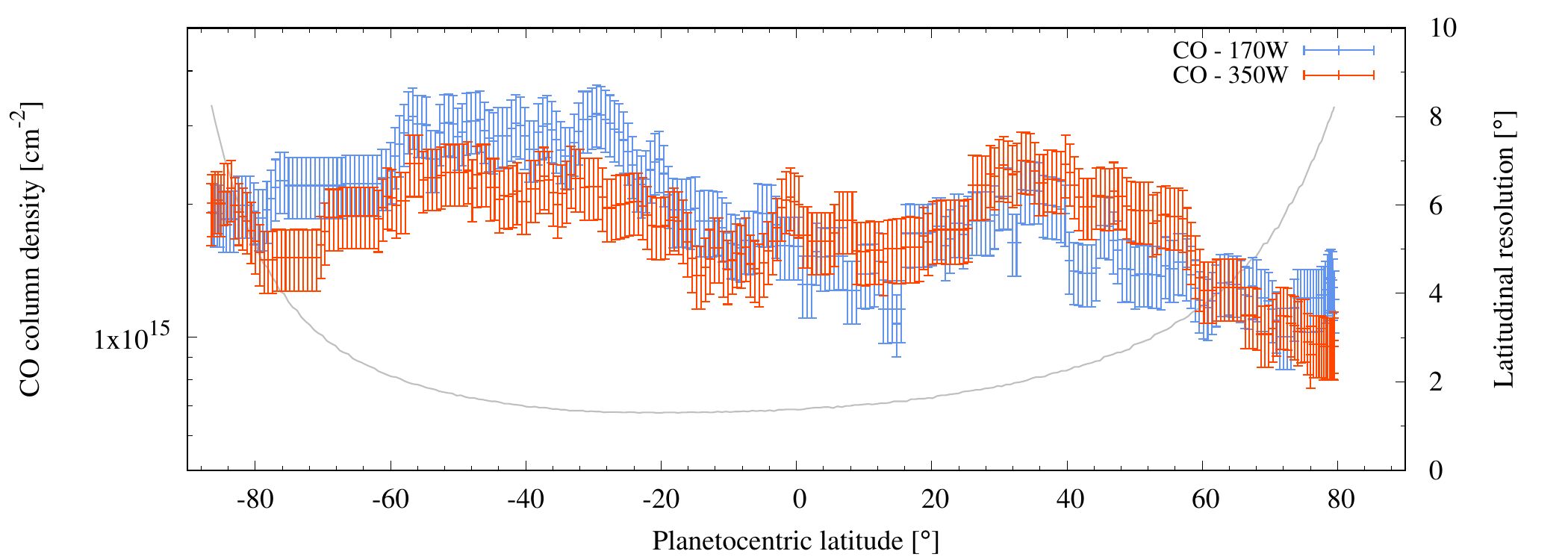}
   \includegraphics[width=15cm,keepaspectratio]{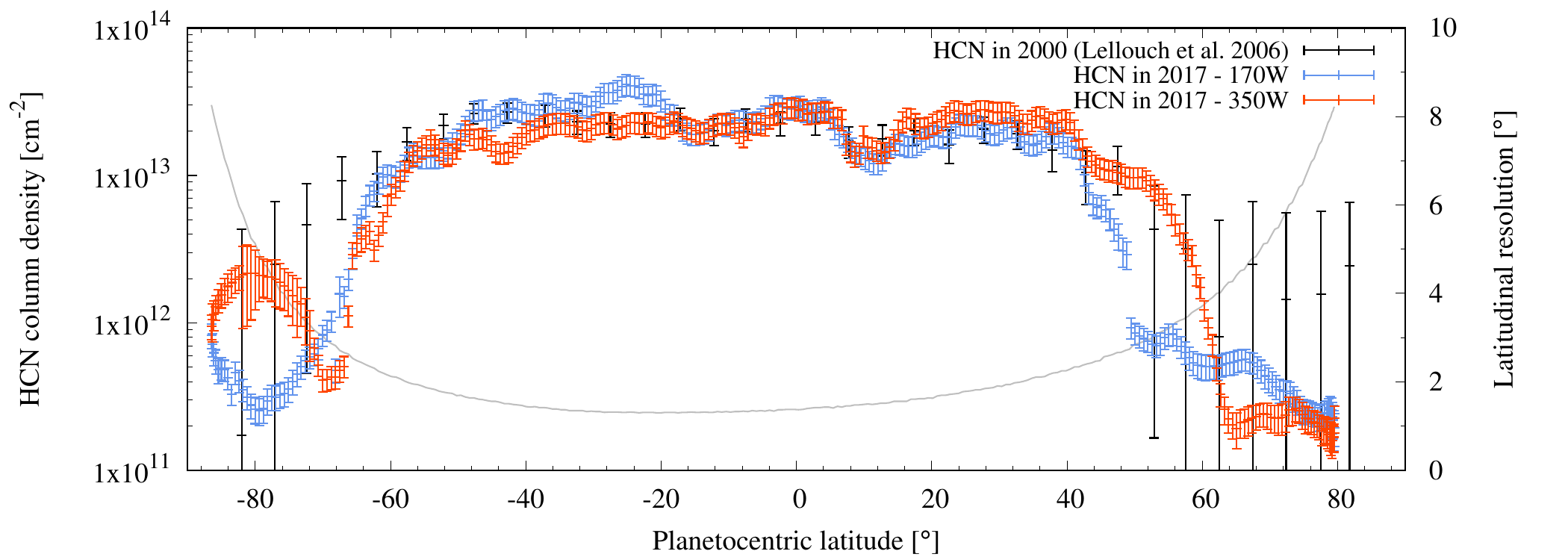}
\end{center}
\caption{CO (top) and HCN (bottom) column densities as a function of planetocentric latitude at the longitudes of the two observed limbs. CO is shown in the top panel and HCN in the bottom one, with mean values and 1-$\sigma$ error bars. The HCN measurements (mean values and 1-$\sigma$ error bars) of (18) are plotted for comparison and the data values have been rescaled by dividing them by 120. The latitudinal resolution of the ALMA observations is shown with the grey line.}
\label{Fig3} 
\end{figure*}

A sample of CO vertical profiles averaged per latitude bins is shown in \fig{Fig2}. The whole set of retrieved profiles is shown in \fig{ExtDataFig3} and we took these profiles to compute the column density as a function of latitude. The meridional distribution of CO is displayed in \fig{Fig3}. It essentially shows that CO was rather uniformly mixed as a function of latitude in Jupiter's stratosphere as of March 2017. We can then put a lower limit on meridional mixing. For CO to populate all latitudes starting from the impact latitude in maximum $\Delta$t$=$ 22.5 years, it requires $K_{yy}\sim\frac{L^{2}}{\Delta t}\geq3.7 \times 10^{11}$ cm$^2$.s$^{-1}$ around mbar pressures, with L the distance from 44$^\circ$S to the north pole. This is consistent with the values derived by (13, 15, 18). With the 3-parameter fit profile, we find a CO mole fraction of 41$\pm$12 ppb (1-$\sigma$ uncertainty) at pressures lower than an average cut-off of 3$\pm$1 mbar. The resulting meridionally uniform CO column density is 1.86$\pm$0.52 $\times$ 10$^{15}$ cm$^{-2}$. The total mass of SL9-derived CO is then 5.47$\pm$0.26 $\times$ 10$^{13}$ g, which corresponds to a loss factor of 9$\pm$3 since 1995-1998 (15). 

The vertical distribution of HCN shows significant variability as seen in \fig{Fig2}. And surprisingly, this variability does not only depend on latitude, as already known from post-SL9 collision observations (17, 18), but also on longitude in the polar domains as demonstrated here by the column density meridional profiles on each limb (\fig{Fig3}). It is rather uniform from $\sim$55$^\circ$S to $\sim$40$^\circ$N with a value of 22.6$\pm$5.7 $\times$ 10$^{12}$ cm$^{-2}$, which is two orders of magnitude lower than that measured 6.5 years after the SL9 impacts by (18). In this latitudinal range, we find an HCN mole fraction of 0.9$\pm$0.5 ppb (1-$\sigma$ uncertainty) at pressures lower than an average cut-off of 2$_{-1}^{+2}$ mbar. The variability seen in the HCN column density, not only on each limb, but also between the eastern and western limbs (at 350W and 170W longitude, respectively), being caused by the continuum subtraction performed at the data reduction stage. It is also true for CO at all latitudes. This subtraction is performed using a uniform disk model which does not perfectly fit the zone/belt structure of Jupiter's continuum. The line-to-continuum ratio (and thus the retrieved value of the HCN mole fraction at high altitude) is consequently altered up to 30\%. On the other hand, the pressure level remains unaffected. We find a similar structure in the variability of the CO column density profile in this latitudinal range and it is thus also not a real feature. At latitudes southward of $\sim$70$^\circ$S, northward of 50$^\circ$N on the 170W limb, and northward of 65$^\circ$N on the 350W limb, the HCN column density drops by a factor of 25--100 with respect to the low-to-mid latitudes. This drop is caused by a depletion of HCN at millibar pressures: the HCN profile is cut-off at a pressure higher than 0.04$_{-0.03}^{+0.07}$ mbar in these regions, translating into much narrower lines compared to those seen in the low-to-mid latitudes (see \fig{Fig1}). In addition, there is an asymmetry in column density between the 350W and 170W limbs in the 45$^\circ$N-65$^\circ$N latitude range. Also, on the 350W limb, between 75$^\circ$S and 85$^\circ$S, the HCN abundance above the cut-off altitude is significantly higher than at other southern polar latitudes. We will discuss these asymmetries in the next two sections. Integrating the HCN column density meridional profile, we find a total mass of 5.0$\pm$0.1 $\times$ 10$^{11}$ g. This corresponds to a loss factor of 50$\pm$30 when compared with the 1995-1998 period (15), or 120$\pm$35 when compared with the value derived in 2000 from the Cassini flyby data (18). The HCN column density was indeed higher in 2000 compared to 1995-1998, probably resulting from secondary production of HCN from NH$_3$ following the comet impacts (18).

\section{HCN at high latitudes}
\begin{figure*}[!h]
\begin{center}
   \includegraphics[width=8cm,keepaspectratio]{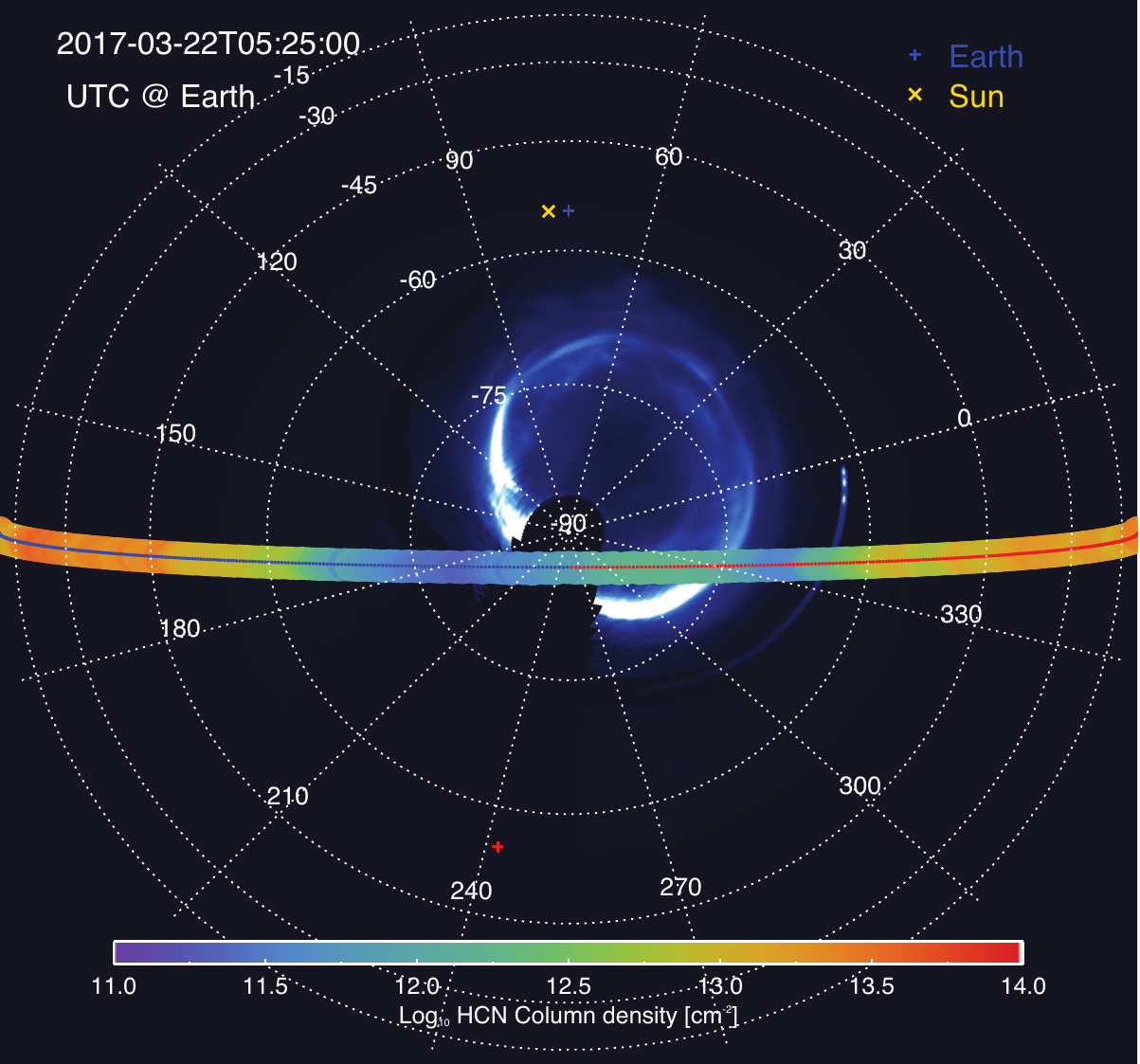}
   \includegraphics[width=8cm,keepaspectratio]{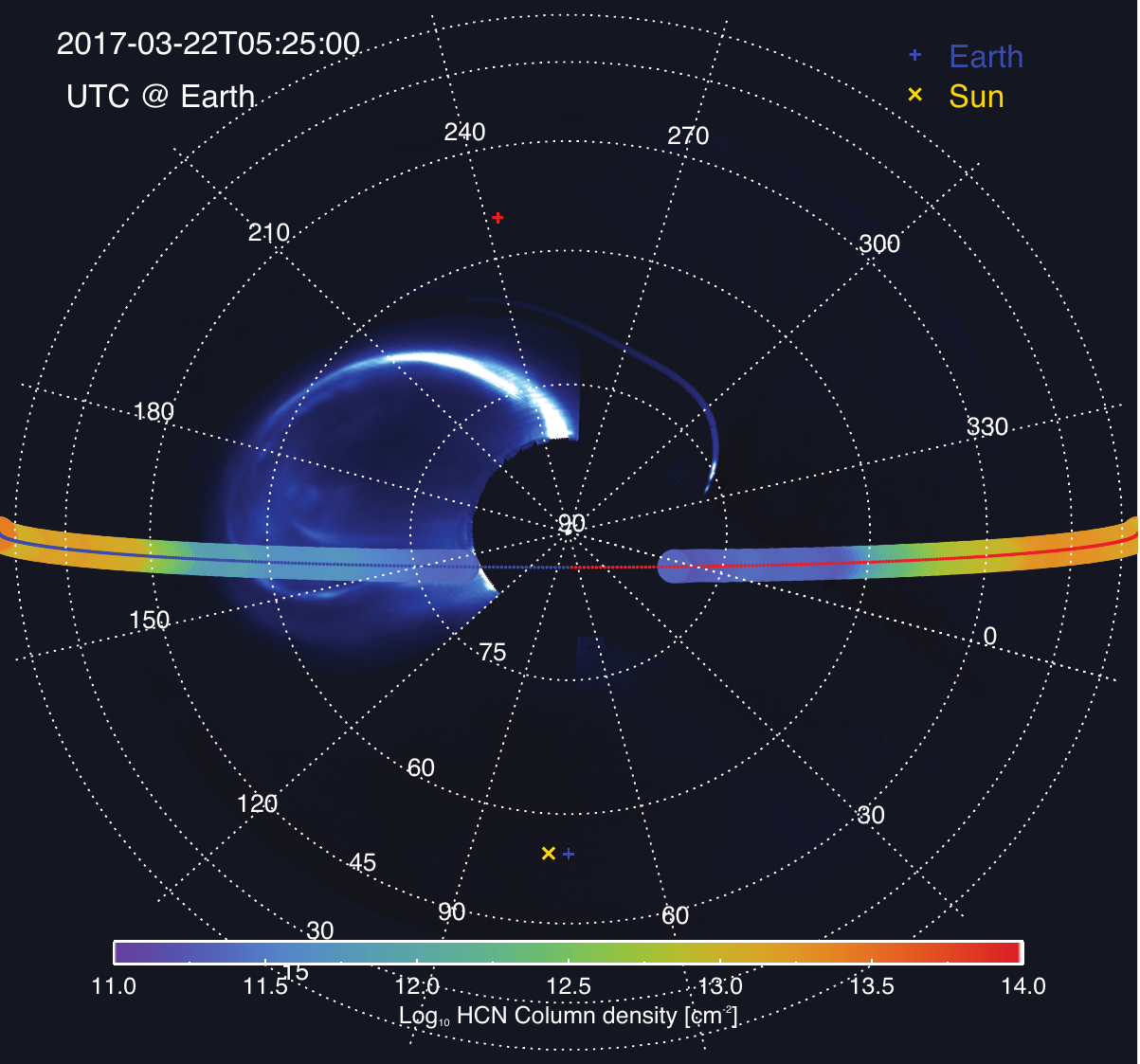}
\end{center}
\caption{Polar projections of the HCN column density. The decimal logarithm of the HCN column density in cm$^{-2}$ is represented according to the observation geometry with ALMA. The solid line in the middle of the column density strip is blue for the western limb and red for the eastern limb. The two panels show the correlation between the HCN depleted region and location of the statistical average of the UV emission of Jupiter's auroras (24). While the southern aurora (left panel) was clearly in the field, the northern aurora (right panel) was rising on the northwest limb. The former highlights the variability of the HCN column density over southern polar and auroral regions. The model of (24) used observations from multiple HST visits. Because of the observation geometrical constraints from HST, not all Jovian longitudes were covered, which resulted in the longitudinal gap seen in both panels. The latitudinal resolution of the ALMA measurements ranges from 4$^\circ$ to 8$^\circ$ northward of 60$^\circ$N along the limb.}
\label{Fig4} 
\end{figure*}

In \fig{Fig4}, we compare the column density on the limb with the statistical average of the UV emission of the auroras (24) at the time of our observations, and we find that a first HCN depleted region lies in the latitude domain occupied by the southern auroral region. Although the geometry was less favorable at the time of the observations with the northern auroral oval just rising over the 170W limb, there is a similar correlation between a second HCN depleted region in the North with the northern auroral region. The asymmetry seen between the northwestern and northeastern latitudes at which the HCN column density drops (\fig{Fig3}) is reminiscent of the tilted position of the more extended northern aurora. The northern and southern HCN depletions seem therefore to be related to the auroras. Because CO and HCN reside at similar altitude levels in the low-to-mid latitudes, they must be subject to the same circulation regime and should share the same meridional distribution. The fact that HCN is clearly depleted in the auroral regions between 0.04 and 3 mbar while CO is not can therefore not be caused by a dynamical barrier that would prevent species located in the mid-latitudes to be transported to higher latitudes. Instead, it requires an efficient chemical loss mechanism specific to HCN at the location of the auroras of Jupiter and at pressures of $\sim$0.1 mbar and higher. Energetic electrons are known to precipitate from the magnetosphere down to the upper stratosphere in Jupiter's auroras and they could destroy HCN. However, we dismiss this possibility because magnetospheric electrons do not penetrate down to mbar pressures in the auroras (25, 26). Consequently, they cannot explain the HCN depletion at pressures higher than 0.1 mbar.

A more promising possibility is heterogeneous chemistry with organic aerosols produced in the auroras. For Titan's atmosphere, (27) proposed that the loss of HCN to solid haze material is an important sink of nitrogen. This idea was extended by (28) and later (29), who showed that introducing a sink term for HCN, representing incorporation into the haze, allowed photochemical model results to better match observations of Titan's HCN and hydrocarbons simultaneously. More recently, (30) have demonstrated that HCN bonds efficiently on organic aerosols produced in laboratory experiments under Titan atmospheric conditions when the aerosols reach a given mass and abundance threshold. The conditions under which these experiments were conducted lead to ion-neutral chemistry initiated by a plasma discharge which are comparable to auroral conditions. Besides, enhanced efficiency in C2 hydrocarbon production has been observed in the jovian auroras by (31), as initially predicted by models of (32). These hydrocarbons are precursors for higher order hydrocarbons in Jupiter's auroral region. Growing carbon chains eventually lead to the formation of benzene (observed by e.g., 33) and eventually polycyclic aromatic hydrocarbons (PAHs) in the 10$^{-3}$--10$^{-1}$ mbar pressure range according to (34). Combining the results of the latter chemistry model with an aerosol microphysical model, (35) predicted that, among PAHs, pyrene first condenses when the pressure exceeds 0.1 mbar and serves as condensation nuclei for phenanthrene and naphthalene. The resulting particles then sediment in the auroral regions as they continue to grow, reaching radii of $\sim$ 0.1 $\mu$m, and eventually lead to the formation of larger aerosols between 0.1 and 1 mbar. Our observations demonstrate that HCN is removed from the gas phase at pressures higher than $\sim$0.1 mbar in the auroras where (35) predict $\sim$ 0.1 $\mu$m aerosols start to form. These aerosols would continue to grow to $\mu$m sizes and sediment inside the auroras, eventually spreading out to other latitudes and longitudes where the auroral winds break (between 0.1 and a few mbar, 21) and accumulating in the middle stratosphere where they were mapped by (36). The smaller and higher altitude aerosols which we presume are responsible for the HCN depletion remain to be observed. The fact that HCN is removed not only interior of the main auroral ovals, but also in the surrounding polar latitudes and longitudes, remains to be explained. In the framework of our interpretation, HCN is no longer depleted between 0.04 and 3 mbar outside the high latitude region, because aerosols are located at too high pressures (36).

\section{HCN interior of the southern auroral oval}
A local increase of the column density can be spotted on the 350W limb at the location of the southern aurora: it can be seen in the data shown in red in \fig{Fig3} in the range 75$^\circ$S-85$^\circ$S, which also corresponds to the vertical profile plotted in red in \fig{Fig2} and to the light blue region interior of the aurora in the polar projection of \fig{Fig4} (left). It reflects an increase in the line emission interior of the southern auroral oval, a region where auroral heating at mbar and sub-mbar levels is known to be significant (31, 37). Since our temperature model is inherently limited by the interpolation that we apply in the few gaps in the coverage of the auroral regions the Gemini/TEXES data have, we may be missing the peak in auroral temperatures that could account at least partially for the excess of HCN emission seen interior of the southern oval. However, there is no profile with increased temperatures in the sub-mbar region that enables the reproduction of the observations without invoking a change in the HCN vertical profile. The HCN line is not optically thick and even increasing the sub-mbar temperatures from 170 K to 210 K decreases the sub-mbar HCN mole fraction only by $\sim$20\% and still fails to properly fit the line. There is thus a $\sim$10 times higher abundance of HCN at sub-mbar pressures interior of the southern aurora compared to the surrounding polar region. Interestingly, observations and models have demonstrated that HCN can be formed by ion-neutral chemistry under auroral-like conditions in Titan's N$_2$-CH$_4$ atmosphere (e.g., 38, 39). And (30) found evidence that HCN formed under such conditions does not bond on organic aerosols (as discussed previously) until small primary monomers have coagulated into large 100 nm-sized monomers. We propose that HCN is formed similarly in Jupiter's auroras and that the source of nitrogen is molecular N$_2$ quenched at kbar pressures from the thermochemical equilibrium between NH$_3$ and N$_2$. Thermochemical model results of (40) indicate that N is $\sim$ 4 $\times$ protosolar in the deep atmosphere, which produces $\sim$10$^{-5}$ of N$_2$ in the upper troposphere. This N$_2$ is transported at all latitudes up to its homopause which resides at 0.1-1 $\mu$bar. It is only ionized in the auroras by magnetospheric electrons, which peak in the 10 nbar-0.1 mbar (25, 26), to produce HCN. This excess of HCN only produced in the jovian auroras would not contaminate other regions of the planet. Indeed, it would remain confined inside the aurora by the auroral winds observed by (21) at least down to pressures where it would be removed by the aerosols. It remains to be shown by chemistry models that this scenario is valid and that there is a chemical pathway to produce $\sim$10 ppb of HCN at pressures lower than 10 $\mu$bar in the auroras (and not elsewhere on the planet) essentially from the ionization (and subsequent dissociation) of N$_2$ coming from the interior by energetic electrons precipitating from the jovian magnetosphere.

These results shed new light on the coupling between magnetospheres and atmospheres in giant planets. New disk-resolved observations of CO$_2$ with JWST (41), possibly coordinated with ALMA mapping observations of CO, HCN, and H$_2$O, as well as dedicated chemical modeling of Jupiter's auroral regions would certainly help consolidate our findings.

\section*{Methods}
\subsection*{Observations}
We use the observations of Jupiter, performed with ALMA on March 22nd, 2017, as part of the 2016.1.01235.S project, which have enabled mapping the winds in Jupiter's stratosphere (21). The rather short integration time of 24 minutes limits the longitudinal smearing to 15$^\circ$. The sub-observer latitude of -3$^\circ$ ensures that only the northernmost latitudes were unobserved. 

The planet was mapped with 42 antennas using 39-point mosaic (see Supplementary Figure 1), because of its 43.82'' angular size. With the C-1 compact antenna configuration, the resulting elliptical synthetic beam was 1.2'' (East-West) $\times$ 1'' (North-South) and the Maximum recoverable Scale was $\sim$8''. Jupiter's disk is thus fully resolved, with a latitudinal resolution of 2$^\circ$ at low latitudes and increasing to 8$^\circ$ at the highest latitudes along the limb. The spectral setup simultaneously covered the HCN (4-3) and CO (3-2) lines at 354.505 and 345.796 GHz, respectively, with spectral resolutions of 122 and 488 kHz, respectively. Such a high spectral resolution enables fully resolving the lineshape of the HCN and CO lines, which have full widths at half-maximum in the range of 2--10 MHz. It is thus possible to determine the vertical profile of the observed species, as detailed in Methods -- section 2.2.

\setcounter{figure}{0}
\begin{figure*}[!h]
\renewcommand\figurename{Supplementary Figure }
\begin{center}
   \includegraphics[width=15cm,keepaspectratio]{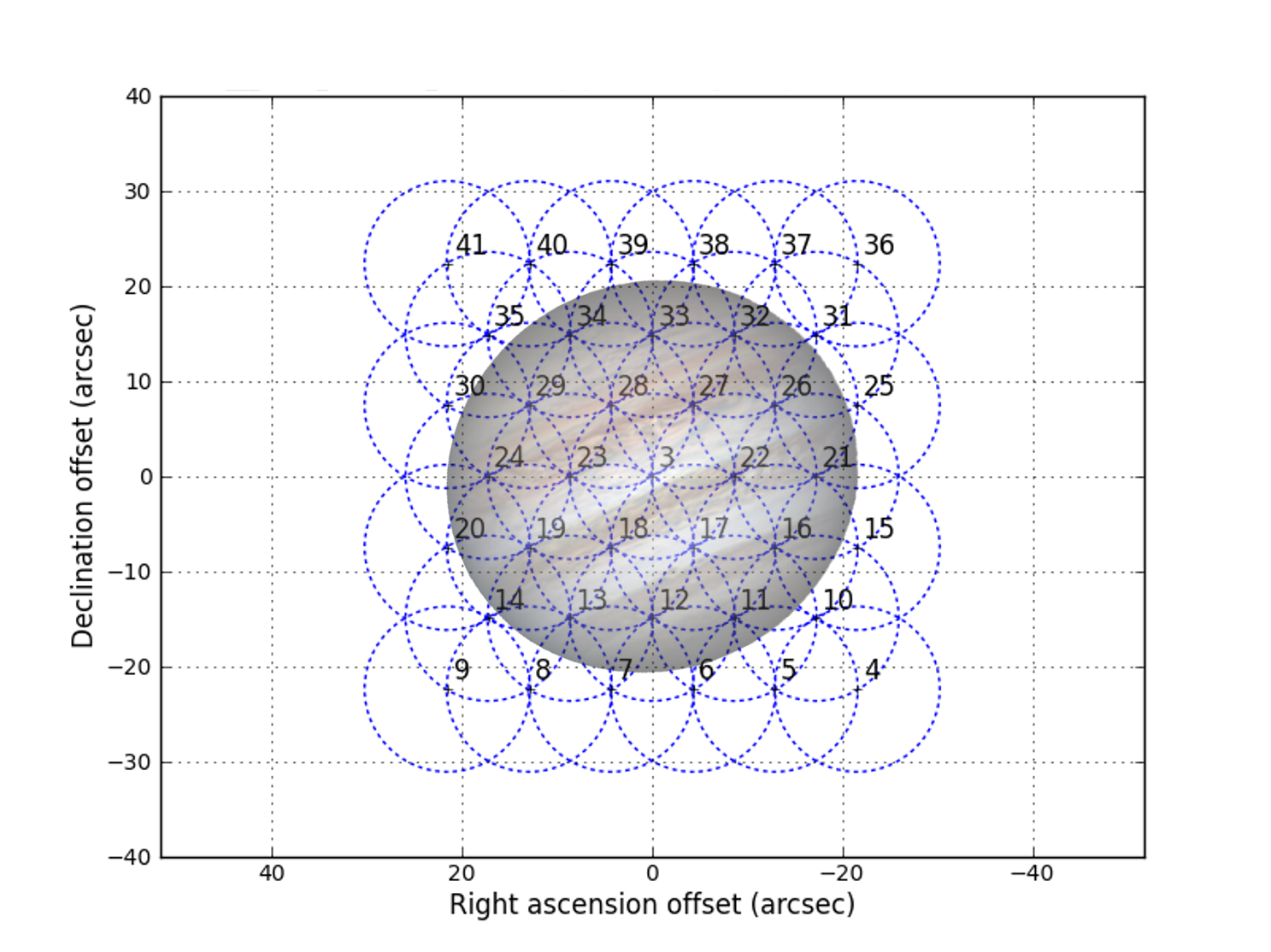}
\end{center}
\caption{Mosaic of 39 points applied to map Jupiter's HCN and CO emission with ALMA on March 22nd, 2017. Jupiter visible image from the same date courtesy of C. Go.}
\label{SuppFig1} 
\end{figure*}

(21) reduced the data using CASA 4.7.2 and applied bandpass, amplitude and phase calibration using the following sources, respectively: J1256-0547 (3C279), Ganymede and J1312-0424. The calibration uncertainties are $\sim$5\%. The noise in the observations is at the level of 3-4\%.

The continuum is then subtracted in the uv-plane to enable imaging the spectral line. It is done using a uniform continuum model. Since we only used the shallow-sounding channels of TEXES we did not retrieve tropospheric temperatures and cannot produce a more realistic continuum map. This results in an imperfect continuum subtraction, which leaves some continuum residuals. These eventually impact the line-to-continuum ratio and the retrieved mole fraction. For example, in a continuum bright region, i.e., with a continuum excess with respect to a uniform disc, the line-to-continuum will be artificially lower and will result in a lower mole fraction. The resulting uncertainty on the column densities is about 30\% and is the biggest source of uncertainty in this study. 

The final Jupiter spectral images show HCN and CO emission concentrated at the planet limb. This is caused by the increase of the path length through the atmosphere and which induces limb-darkening in the continuum and limb-brightening in the spectral line. We took the 557 HCN spectra and 540 CO spectra they extracted from the limb (at the 1-bar level) to carry out our analyses. The beam was oversampled by a factor of four to five and the spectra were obtained from a bilinear interpolation of the spatial pixels of the data cube. Since the angular resolution is slightly better in the HCN data, we extract slightly more spectra from this datacube.

\subsection*{Spatial distribution retrieval method}
We analyzed each CO and HCN limb spectrum with a model that enabled us to retrieve their vertical profile all along the limb. This model is composed of a forward radiative transfer model and an iterative retrieval algorithm relying on a three-parameter function for vertical profile parametrization. We detail these two components hereafter.

\subsubsection{Forward model}
\paragraph{Radiative transfer}
We used the radiative transfer model described in (42) adapted to Jupiter's atmosphere. It accounts for the 3D ellipsoidal geometry of the planet. We take the ephemeris from JPL Horizons Solar System Dynamics ephemerides (https://ssd.jpl.nasa.gov/horizons/) to have the planet geometrical data at the time of the observations. In particular, we use the equatorial angular diameter, the observer sub-latitude and sub-longitude, North Pole angle, and geocentric distance.

The spectroscopic data for both lines come from the JPL catalog. We adopted a pressure broadening coefficient $\gamma$ of 0.145 cm$^{-1}$.atm$^{-1}$ and a temperature dependence exponent $n$ of 0.75 for the HCN line according to (43). For the CO line, we took $\gamma$=0.067 cm$^{-1}$.atm$^{-1}$ and $n$=0.60, according to (44).

The continuum is formed from the collision-induced absorption of H$_2$--He--CH$_4$ (45-47) and the far wings of NH$_3$ and PH$_3$ lines. We parametrize the vertical profile of both species using $q=q_0 (\frac{p}{p_0})^{((1-f)/f)}$ with deep mole fractions $q_0$ of 2$\times$10$^{-4}$ and 6$\times$10$^{-7}$, cut-off pressures $p_0$ of 800 and 500 mbar, and fractional scale heights f of 0.15 and 0.2, for NH$_3$ and for PH$_3$, respectively.

\paragraph{Temperature field}
The temperature vertical profile in the stratosphere of Jupiter shows meridional, and (to a smaller extent) zonal variability over the course of only a few weeks in several regions like the equatorial and auroral regions (e.g., 48, 49). Fortunately, Jupiter's stratospheric temperatures have been retrieved from Gemini/TEXES observations of the CH$_4$ band at 7.8 $\mu$m a few days before our observations. The TEXES observations record high resolution spectra in 4-6 cm$^{-1}$ wide bands centered at 587, 730, 819, 950 and 1248 cm$^{-1}$. The temperatures are retrieved from the 587 cm$^{-1}$ and 1248 cm$^{-1}$ settings, which collectively sound 50 mbar to 1 $\mu$bar, as shown with vertical functional derivatives in Fig. 3 of (31). Upper tropospheric temperatures are only constrained at 100 mbar but not deeper because we did not use the deeper-sounding channels of TEXES in our retrievals. The effect on our abundance retrievals remains insignificant compared to other uncertainty sources. The spectral maps of the equatorial zone up to the mid-latitudes were obtained on March 14, 16 and 20, 2017 by (22) and for the high latitudes on March 17--19, 2017, by (23). The angular resolution of the observations was diffraction-limited to $\sim$0.7'' at 7.8 $\mu$m, where the CH$_4$ emissions were inverted to derive the stratospheric temperature profile, with peak sensitivities between 1 $\mu$bar and 50 mbar. For instance, the angular resolution corresponds to a latitude-longitude footprint of $\sim$5 degrees at 60$^\circ$N, which is of the same order as that of the ALMA observations. We reconstructed a full 3D temperature field from the combined retrieved temperature maps, and filled the latitudinal and longitudinal gaps by interpolating linearly between the data points. This results in some discrete jumps as those seen at 61$^\circ$N and 66$^\circ$S, which nonetheless result from the marked boundaries of auroral heating. To speed up the retrievals, we used 2D latitude/pressure maps instead of the full 3D temperature field. We extracted these 2D thermal maps from the 3D field at 350W (eastern limb) and 170W (western limb). They are displayed in \fig{ExtDataFig1} and account for the 15$^\circ$ longitudinal smearing of the observations. Note that the heating caused by the northern and southern auroras can be seen on the 170W and 350W plots, respectively. Cooling in the troposphere and lower stratosphere poleward of $\sim$60$^\circ$ is also present. It modifies the local continuum, but barely impacts the line amplitude (effect $<$ 5\%). It thus has negligible effects on the retrieved abundances. In a first and simple approach, we do not consider the temperatures at other longitudes in our radiative transfer calculations and essentially work with 2D fields (pressure-latitude) for each limb.

\setcounter{figure}{0}
\begin{figure*}[!h]
\renewcommand\figurename{Extended Data Figure }
\begin{center}
   \includegraphics[width=8cm,keepaspectratio]{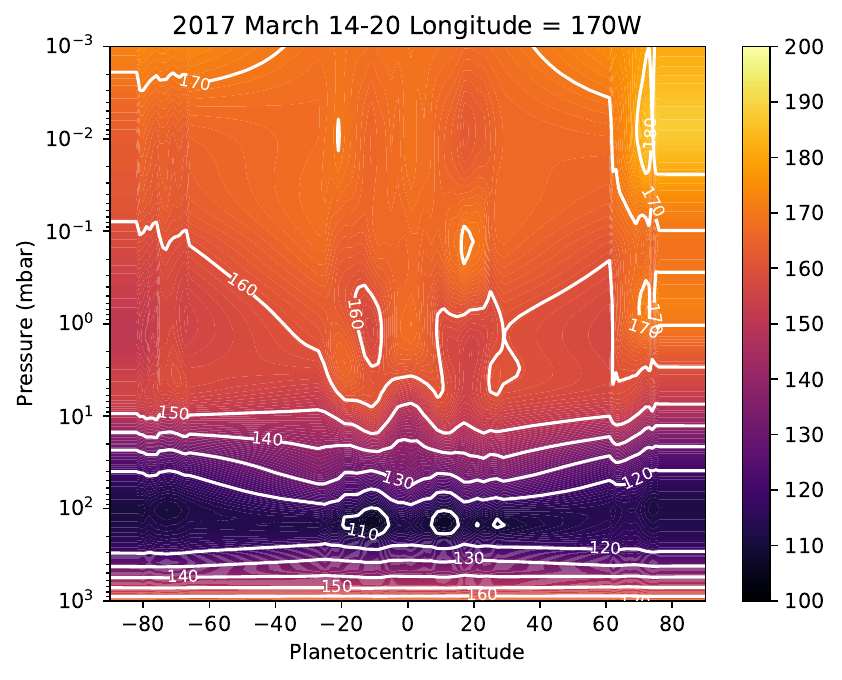}
   \includegraphics[width=8cm,keepaspectratio]{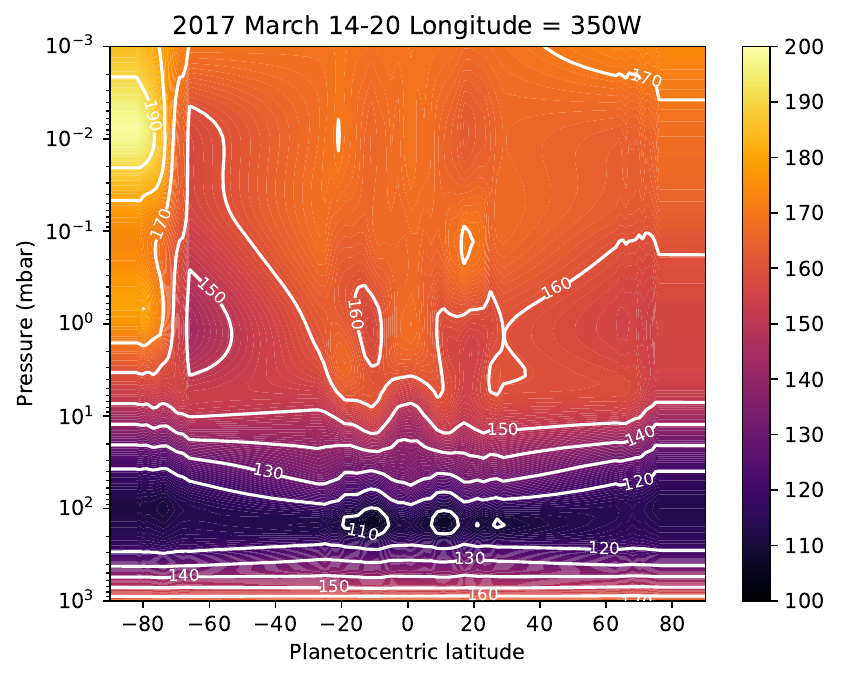}
\end{center}
\caption{Temperature fields used in the abundance retrievals. The fields at 170W (left) and 350W (right) correspond to the eastern and western limbs, respectively, as reconstructed from the retrievals obtained from Gemini/TEXES observations on March 14--20, 2017. }
\label{ExtDataFig1} 
\end{figure*}

\subsubsection{Retrieval model}
With known temperature and pressure fields, we used the ALMA observations to obtain vertical profiles of volume mixing ratio of HCN and CO. Using a simple two-parameter profile parametrization, with mixing ratio above a pressure cut-off level, would be insufficient to properly fit the HCN lines, especially in the near-continuum wings of the line. We thus opted for a three-parameter parametrization of the vertical profiles, because the HCN line profile clearly implies a cut-off altitude which has to be described by two parameters (slope of decay and altitude where the decay begins). The mixing ratio as a function of height in km above the 1 bar level, f(z), is parameterized with three independent numbers, $a_1$, $a_2$, $a_3$ as follows:
\begin{equation}
f(z)=a_1 (1+\tanh \left(z-a_2\right)/a_3 )
\end{equation}
The inverse problem is then to find only three numbers at each of the limb spectra, where $a_1$ is the high altitude asymptotical constant mixing ratio above a transition level, $a_2$ is the altitude of this transition level, and $a_3$ is a scale height informing how quickly the mixing ratio decreases below the transition level with decreasing altitude.

\begin{figure*}[!h]
\renewcommand\figurename{Extended Data Figure }
\begin{center}
   \includegraphics[width=8cm,keepaspectratio]{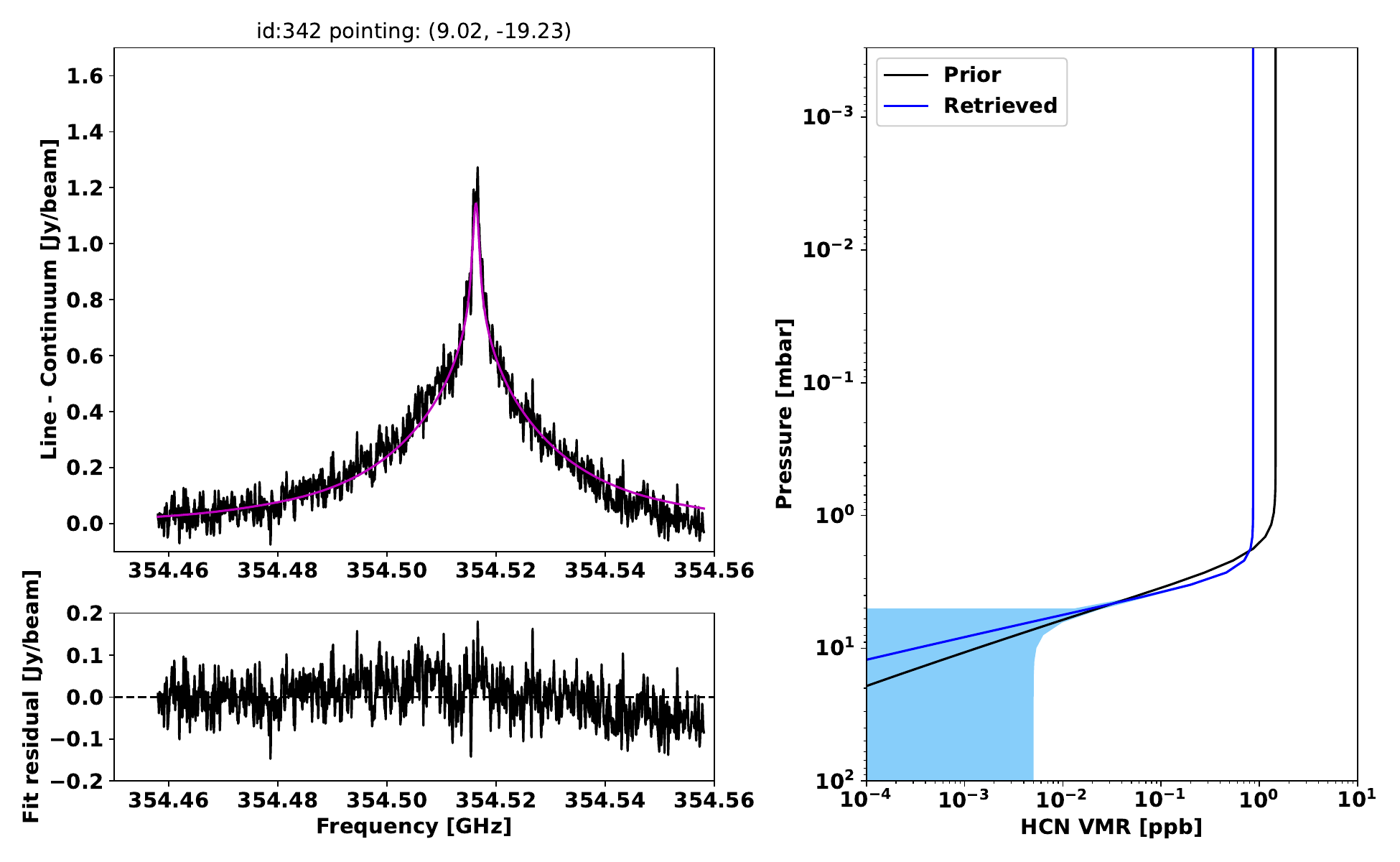}
   \includegraphics[width=8cm,keepaspectratio]{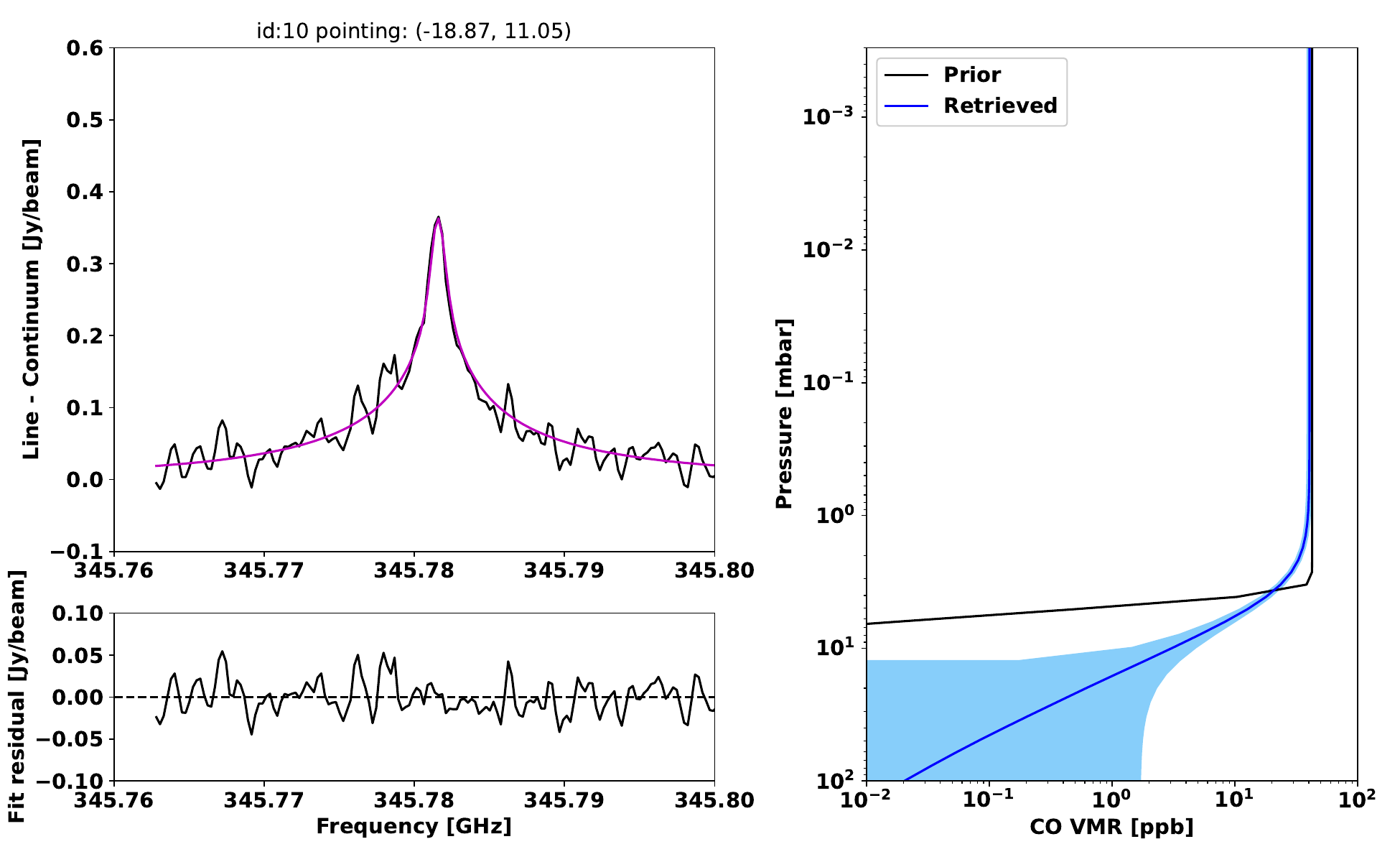}
\end{center}
\caption{HCN (left) and CO (right) VMR vertical profile retrieval examples. The a priori and retrieved profiles are shown in black and blue, respectively. The shaded region encompasses the range of 1-$\sigma$ uncertainties due to random measurement errors. The measurement spectra (black) and fitted spectra (magenta) along with the residuals are shown as well.}
\label{ExtDataFig2} 
\end{figure*}

The choice of this parametrization provides several advantages worth pointing out. With only three unknown the inverse problem is robust and stable, it is relatively fast, and it does yield already a high quality of fitting (see \fig{ExtDataFig2}), indicating that we exploit most of the information content of the spectra. In addition, the parameterized vertical profiles capture the key physical behavior of the actual HCN and CO mixing ratios with cut-off height as a free parameter. We studied the forward radiance Jacobians (as used in inverse problems) and found that the three parameters can be estimated with a high-degree of independence. Also, the stability of the inverse problem implies no degeneracy between the three parameters.

We solved the inverse problem with an iterative damped least-squares method, also known as Twomey--Tikhonov in Earth atmospheric science literature (50). The three parameters are assembled into a vector $\mathbf{x}$, and the next iteration is obtained as
\begin{equation}
\mathbf{x_{i+1}}=\mathbf{x_i} + (\mathbf{K}^{T} \mathbf{S_e}^{-1} \mathbf{K} + \mathbf{R})^-{1} (\mathbf{K}^{T} \mathbf{S_e}^{-1} \mathbf{\Delta y} - \mathbf{Rx}_{i}).
\end{equation}
The Jacobian matrix $\mathbf{K}=\frac{\partial y_{n}}{\partial x_{m}}$, where $\mathbf{y_n}$ is the n-th spectral channel of the forward model, was estimated numerically perturbing each of the three parameters by a small amount and recalculating forward model spectra. This matrix remains constant during iterations. The diagonal matrices $\mathbf{S_{e}}$ and $\mathbf{R}=\alpha\mathbf{I}$ denote random measurement noise (1-$\sigma$) and a regularization operator respectively. The regularization parameter, $\alpha$, was estimated by trial and error with synthetic inversions and then fixed for all subsequent iterations of all retrievals. The $\mathbf{\Delta y}$ term holds differences between measured and calculated spectra for the current estimate of $\mathbf{x}$. The iterative process continues while the current reduced $\chi_i^2$ is smaller than the one from previous iteration within a threshold of 0.5\%.

\begin{figure*}[!h]
\renewcommand\figurename{Extended Data Figure }
\begin{center}
   \includegraphics[width=8cm,keepaspectratio]{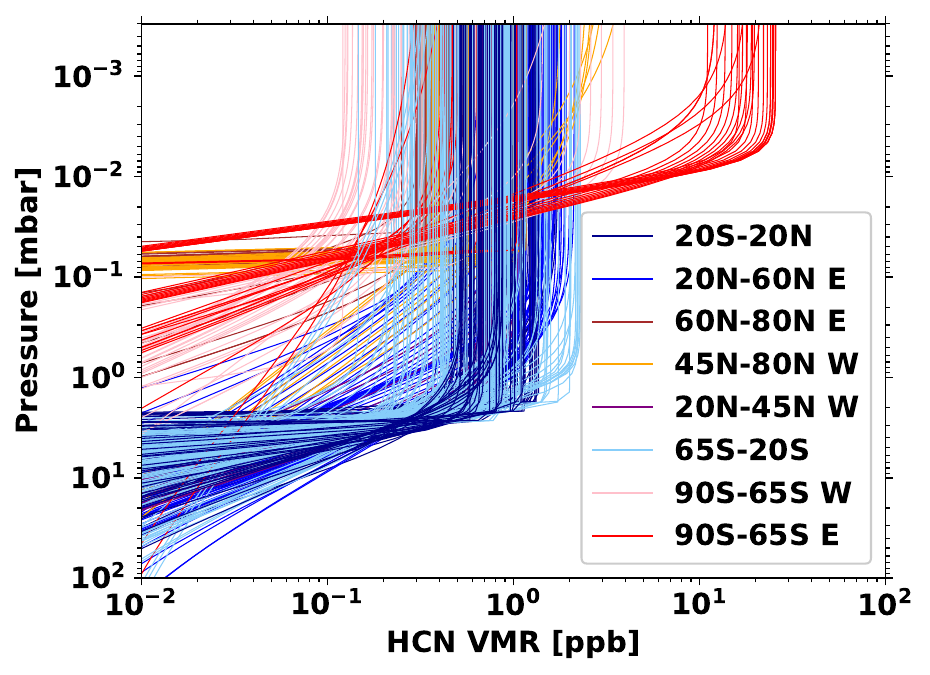}
   \includegraphics[width=8cm,keepaspectratio]{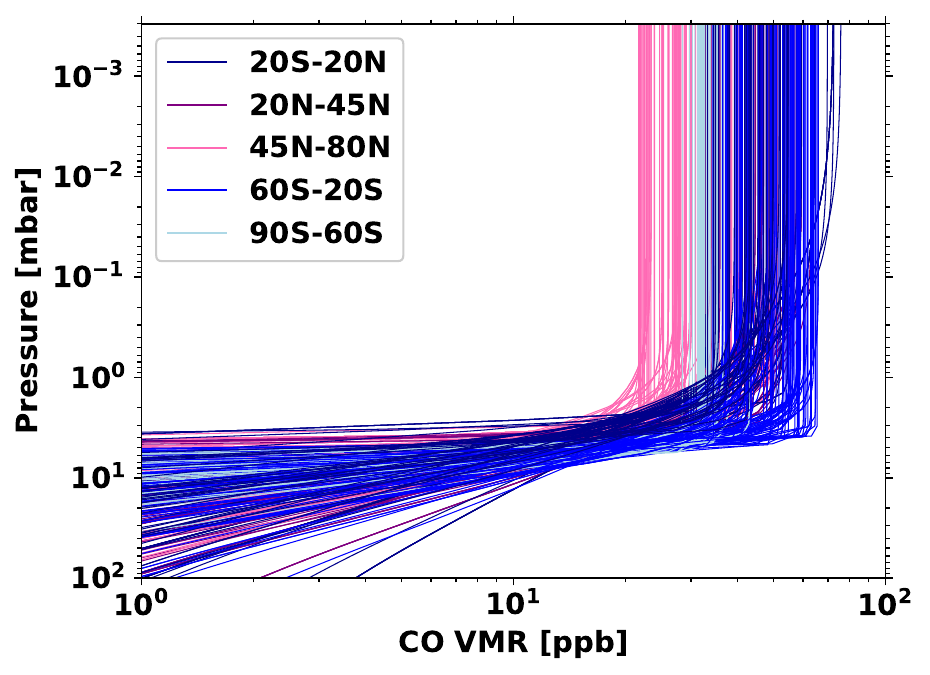}
\end{center}
\caption{HCN (left) and CO (right) vertical profiles in Jupiter's stratosphere, as retrieved from the ALMA observations of March 22nd, 2017. They are grouped in latitude bins, similarly to \fig{Fig2}.}
\label{ExtDataFig3} 
\end{figure*}

The retrieval process described above requires an initial estimate on the three parameter values for each new observation. For the first inverted spectrum, these were supplied manually based on a few trials. For subsequent points on the map, they were however taken from the previously inverted measurement before running the first retrieval iteration. This approach is reasonable as we expect adjacent points on the map not to deviate strongly from one another (in mixing ratio). For validation of our assumption, we investigated the spectral line characteristics of adjacent observations. Finally, the least-squares method allowed us to estimate how the random error component in the measurements gets projected into uncertainties in retrieved parameters through the solution covariance matrix. The diagonal elements were used in plotting the retrieved profile uncertainties (as shaded regions in \fig{ExtDataFig2}) which are generally smaller than 3\% on the $a_1$ parameter, smaller than 1\% on the $a_2$ parameter, and in the 5-10\% range on the $a_3$ parameter.
Random samples of retrieved HCN and CO profiles with the corresponding spectra and fit are displayed in \fig{ExtDataFig2}. About 95\% of HCN retrievals converge automatically to the set criteria, and about 98\% in case of CO which is more optically thin. In the few cases which did not automatically converge, we supplied manually a better initial value of the $a_2$ parameter which allowed the inverse problem to proceed nominally to convergence. The full set of HCN and CO vertical profiles, as retrieved from the data, is displayed in \fig{ExtDataFig3} using the same color code as in \fig{Fig2} for the various latitude bins.

In addition, we investigated the role of the full 3D temperature field on the mixing ratio retrievals, especially at auroral latitudes where sharp gradients may exist. We compared forward radiative calculations of the limb emission using the retrieved volume mixing ratios and line-of-sights going through the full 3D temperature field, with calculations using the same abundance profiles and the 2D temperature maps produced for the two limbs (\fig{ExtDataFig1}). In the majority of cases, the 3D temperatures did not play any role. However, in a few cases, we identified a need to consider the full 3D inputs in the inversion of the HCN profiles. In these limited cases where the 2D temperatures would produce a fit outside the measurement noise, we rerun the retrieval procedure with the full 3D temperature field. These cases represent less than 5\% of the HCN spectra and are located at 33$^\circ$S--46$^\circ$S and 53$^\circ$S--57$^\circ$S on the 350W limb and 64$^\circ$S--69$^\circ$S on the 170W limb.

\section*{Data availability}
Observation data can be obtained from the ALMA archive. Temperature profiles, HCN and CO limb spectra and retrieved vertical profiles can be obtained on the following Zenodo repository: https://doi.org/10.5281/zenodo.7928701

\section*{Code availability}
Software used for the retrievals is available upon request by contacting the corresponding author.

\section*{Acknowledgements}
T.C. acknowledges funding from CNES and the Programme National de Plan\'etologie (PNP) of CNRS/INSU. 
This paper makes use of the following ALMA data: ADS/JAO.ALMA\#2016.1.01235.S. ALMA is a partnership of ESO (representing its member states), NSF (USA) and NINS (Japan), together with NRC (Canada), MOST and ASIAA (Taiwan), and KASI (Republic of Korea), in cooperation with the Republic of Chile. The Joint ALMA Observatory is operated by ESO, AUI/NRAO and NAOJ.

\section*{Author contribution statement}
T.C. and L. R. performed the modeling and data analysis. All authors discussed the results and commented on the manuscript. 

\section*{Competing interest statement}
Authors declare that they have no competing interests.


\section*{References}
\begin{enumerate}
  \item Moses, J. I., Fouchet, T., B\'ezard, B., Gladstone, G. R., Lellouch, E., Feuchtgruber, H. Photochemistry and diffusion in Jupiter?s stratosphere: Constraints from ISO observations and comparisons with other planets. J. Geophys. Res. 110, E08001 (2005).
  \item Feuchtgruber, H., Lellouch, E., de Graauw, T., B\'ezard, B., Encrenaz, T., Griffin, M. External supply of oxygen to the atmospheres of the giant planets. Nature 389, 159-162 (1997).
  \item Hue, V., Hersant, F., Cavali\'e, T., Dobrijevic, M., Sinclair, J. A. Photochemistry, mixing and transport in Jupiter?s stratosphere constrained by Cassini. Icarus 307, 106-123 (2018).
  \item Harrington, J., de Pater, I., Brecht, S. H., Drake, D., Meadows, V., Zahnle, K., Nicholson, P. D. Lessons from Shoemaker-Levy 9 about Jupiter and planetary impacts. In: Jupiter. The planet, satellites and magnetosphere. Edited by Fran Bagenal, Timothy E. Dowling, William B. McKinnon. Cambridge planetary science, Vol. 1, Cambridge, UK: Cambridge University Press, ISBN 0-521-81808-7, 159 ? 184 (2004).
  \item Orton, G. A?Hearn, M., Baines, K., Deming, D., Dowling, T., Goguen, J., Griffith, C., Hammel, H., Hoffmann, W., Hunten, D., Jewitt, D., Kostiuk, T., Miller, S., Noll, K., Zahnle, K., Achilleos, N., Dayal, A., Deutsch, L., Espenak, F., Esterle, P., Friedson, J., Fast, K., Harrington, J., Hora, J., Joseph, R., Kelly, D., Knacke, R., Lacy, J., Lisse, C., Rayner, J., Sprague, A., Shure, A., Wells, K., Yanamandra-Fisher, P., Zipoy, D., Bjoraker, G., Buhi, D., Golisch, W., Griep, D., Kaminski, C., Arden, C., Chaikin, A., Goldstein, J., Gilmore, D., Fazio, G., Kanamori, T., Lam, H., Livengood, T., Maclow, M.-M;, Marley, M., Momary, T., Robertson, D., Romani, P., Spitale, J., Sykes, M., Tennyson, J., Wellnitz, D., Ying, S.-W. Collision of comet Shoemaker-Levy 9 with Jupiter observed by the NASA Infrared Telescope Facility. Science 267, 1277-1282 (1995).
  \item Moreno, R., Marten, A., Biraud, Y., B\'ezard, B., Lellouch, E., Paubert, G., Wild, W. Jovian stratospheric temperature during the two months following the impacts of comet Shoemaker-Levy 9. Planet. Space Sci. 49, 473-486 (2001).
  \item S\'anchez-Lavega, A., Lecacheux, J., Colas, F., Gomez, J. M., Laques, P., Miyazaki, I., Parker, D. C. Motions of the SL9 impact clouds. Geophys. Res. Lett. 22, 1761-1764 (1995).
  \item Hammel, H. B., Beebe, R. F., Ingersoll, A. P., Orton, G. S., Mills, J. R., Simon, A. A., Chodas, P., Clarke, J. T., de Jong, E., Dowling, T. E., Harrington, J., Huber, L. F., Karkoschka, E., Santori, C. M., Toigo, A., Yeomans, D., West, R. A. HST imaging of atmospheric phenomena created by the impact of comet Shoemaker-Levy9. Science 267, 1288-1296 (1995).
  \item Lellouch, E., Paubert, G., Moreno, R., Festou, M. C., B\'ezard, B., Bockel\'ee-Morvan, D., Colom, P., Crovisier, J., Encrenaz, T., Gautier, D., Marten, A., Despois, D., Strobel, D. F., Sievers, A. Chemical response of Jupiter?s atmosphere following the impact of comet Shoemaker-Levy 9. Nature 373, 592-595 (1995).
  \item Marten, A., Gautier, D., Griffin, M. J., Matthews, H. E., Naylor, D. A., Davis, G. R., Owen, T., Orton, G., Bockel\'ee-Morvan, D., Colom, P., Crovisier, J., Lellouch, E., de Pater, I., Atreya, S., Strobel, D., Han, B., Sanders, D. B. The collision of comet Shoemaker-Levy 9 with Jupiter: Detection and evolution of HCN in the stratosphere of the planet. Geophys. Res. Lett. 22, 1589-1592 (1995).
  \item Bjoraker, G. L., Stolovy, S. R., Herter, T. L., Gull, G. E., Pirger, B. E. Detection of water after the collision of fragments G and K of comet Shoemaker-Levy 9 with Jupiter. Icarus 121, 411-421 (1996).
  \item Zahnle, K. Dynamics and chemistry of SL9 plumes. In IAU Colloq. 156: The Collision of Comet Shoemaker-Levy 9 and Jupiter, ed. K. S. Noll, H. A. Weaver, \& P.D. Feldman, 183-212 (1996).
  \item Lellouch, E., B\'ezard, B., Moses, J. I., Davis, G. R., Drossart, P., Feuchtgruber, H., Bergin, E. A., Moreno, R., Encrenaz, T. The origin of water vapor and carbon dioxide in Jupiter?s stratosphere. Icarus 159, 112-131 (2002).
  \item Lellouch, E., B\'ezard, B., Moreno, R., Bockel\'ee-Morvan, D., Colom, P., Festou, M., Gautier, D., Marten, A., Paubert, G. Carbon monoxide in Jupiter after the impact of comet Shoemaker-Levy 9. Planet. Space Sci. 45, 1203-1212 (1997).
  \item Moreno, R., Marten, A., Matthews, H. E., Biraud, Y. Long-term evolution of CO, CS and HCN in Jupiter after the impacts of comet Shoemaker-Levy 9. Planet. Space Sci. 51, 591-611 (2003).
  \item Griffith, C. A., B\'ezard, B., Greathouse, T. K., Kelly, D. M., Lacy, J. H., Noll, K. S. Thermal infrared imaging spectroscopy of Shoemaker-Levy 9 impact sites: spatial and vertical distributions of NH3, C2H4, and 10-$\mu$m dust emission. Icarus 128, 275-293 (1997).
  \item Griffith, C. A., B\'ezard, B., Greathouse, T., Lellouch, E., Lacy, J., Kelly, D., Richter, M. J. Meridional transport of HCN from SL9 impacts on Jupiter. Icarus 170, 58-69 (2004).
  \item Lellouch, E., B\'ezard, B., Strobel, D. F., Bjoraker, G. L., Flasar, F. M., Romani, P. N. On the HCN and CO2 abundance and distribution in Jupiter?s stratosphere. Icarus 184, 478-497 (2006).
  \item Cavali\'e, T., Billebaud, F., Biver, N., Dobrijevic, M., Lellouch, E., Brillet, J., Lecacheux, A., Hjalmarson, \r{A}, Sandqvist, Aa., Frisk, U., Olberg, M., Bergin, E. A., The Odin Team. Observation of water vapor in the stratosphere of Jupiter with the Odin space telescope. Planet. Space Sci. 56, 1573-1584 (2008).
  \item Benmahi, B., Cavali\'e, T., Dobrijevic, M., Biver, N., Bermudez-Diaz, K., Aa. Sandqvist, Lellouch, E., Moreno, R., Fouchet, T., Hue, V., Hartogh, P., Billebaud, F., Lecacheux, A., Hjalmarson, \r{A}, Frisk, U., Olberg, M., The Odin Team. Monitoring of the evolution of H$_2$O vapor in the stratosphere of Jupiter over an 18-year period with the Odin space telescope. Astron. Astrophys. 641, A140 (2020).
  \item Cavali\'e, T., Benmahi, B., Hue, V., Moreno, R., Lellouch, E., Fouchet, T., Hartogh, P., Rezac, L., Greathouse, T. K., Gladstone, G. R., Sinclair, J. A., Dobrijevic, M., Billebaud, F., Jarchow, C. First direct measurement of auroral and equatorial jets in the stratosphere of Jupiter. Astron. Astrophys. 647, L8 (2021).
  \item Cosentino, R. G., Morales-Juber\'ias, R., Greathouse, T., Orton, G., Johnson, P., Fletcher, L. N., Simon, A. New observations and modeling of Jupiter?s quasi-quadrennial oscillation. J. Geophys. Res. Planets 122, 2719-2744.
  \item Sinclair, J. A., Greathouse, T. K., Giles, R. S., Lacy, J., Moses, J., Hue, V., Grodent, D., Bonfond, B., Tao, C., Cavali\'e, T., Dahl, E. K., Orton, G. S., Fletcher, L. N., Irwin P. G. J. A high spatial and spectral resolution study of Jupiter?s mid-infrared auroral emissions during a solar wind compression. Planet. Sci. J. 4, 76 (2023).
  \item Clarke, J. T., Nichols, J., G\'erard, J.-C., Grodent, D., Hansen, K. C., Kurth, W., Gladstone, G. R., Duval, J., Wannawichian, S., Bunce, E., Cowley, S. W. H., Crary, F., Dougherty, M., Lamy, L., Mitchell, D., Pryor, W., Retherford, K., Stallard, T., Zieger, B., Zarka, P., Cecconi, B. Response of Jupiter?s and Saturn?s auroral activity to the solar wind. J. Geophys. Res. 114, A05210 (2009).
  \item Perry, J. J., Kim, Y. H., Fox, J. L., Porter, H. S. Chemistry of the jovian auroral ionosphere. J. Geophys. Res. 104, 16541-16565 (1999).
  \item G\'erard, J.-C., Bonfond, B., Grodent, D., Radioti, A., Clarke, J. T., Gladstone, G. R., Waite, J. H., Bisikalo, D., Shematovich, V. I. Mapping the electorn energy in Jupiter?s aurora: Hubble spectral observations. J. Geophys. Res. 119, 9072-9088 (2014).
  \item McKay, C. P. Elemental composition, solubility, and optical properties of Titan?s organic haze. Planet. Space Sci. 44, 741-747 (1996).
  \item Lara, L.-M., Lellouch, E., Shematovich, V. Titan?s atmospheric haze: the case for HCN incorporation. Astron. Astrophys. 341, 312-317 (1999).
  \item Vinatier, S., B\'ezard, B., Fouchet, T., Teanby, N. A., de Kok, R., Irwin, P. G. J., Conrath, B. J., Nixon, C. A., Romani, P. N., Flasar, F. M., Coustenis A. Vertical abundance profiles of hydrocarbons in Titan?s atmosphere at 15$^\circ$S and 80$^\circ$N retrieved from Cassini/CIRS spectra. Icarus 188, 120-138 (2007).
  \item Perrin, Z., Carrasco, N., Chatain, A., Jovanovic, L., Vettier, L., Ruscassier, N., Cernogora, G. An atmospheric origin for HCN-derived polymers on Titan. Processes 9, 965 (2021).
  \item Sinclair, J. A., Orton, G. S., Greathouse, T. K., Fletcher, L. N., Moses, J. I., Hue, V., Irwin, P. G. J. Jupiter?s auroral-related stratospheric heating and chemistry II: Analysis of IRTF-TEXES spectra measured in December 2004. Icarus 300, 305-326 (2018).
  \item Wong, A.-S., Yung, Y. L., Friedson, A. J. Benzene and haze formation in the polar atmosphere of Jupiter. Geophys. Res. Lett. 30, 1447 (2003).
  \item B\'ezard, B., Drossart, P., Encrenaz, T., Feuchtgruber, H. Benzene on the giant planets. Icarus 154, 492-500 (2001).
  \item Wong, A.-S., Lee, A. Y. T., Yung, Y. L., Ajello, J. M. Jupiter: aerosol chemistry in the polar atmosphere. Astrophys. J. 534, L215-L217 (2000).
  \item Friedson, A. J., Wong, A.-S., Yung, Y. L. Models for polar haze formation in Jupiter?s stratosphere. Icarus 158, 389-400 (2002).
  \item Zhang, X., West, R. A., Banfield, D., Yung, Y. L. Stratospheric aerosols on Jupiter from Cassini observations. Icarus 226, 159-171 (2013).
  \item Sinclair, J. A., Orton, G. S., Greathouse, T. K., Fletcher, L. N., Moses, J. I., Hue, V., Irwin, P.G.J. Jupiter?s auroral-related stratospheric heating and chemistry I: Analysis of Voyager-IRIS and Cassini-CIRS spectra. Icarus 292, 182-207 (2017).
  \item Waite, J. H., Young, D. T., Cravens, T. E., Coates, A. J., Crary, F. J., Magee, B., Westlake, J. The process of tholin formation in Titan?s upper atmosphere. Science 316, 870-875 (2007).
  \item Dobrijevic, M., Loison, J.-C., Hickson, K. M., Gronoff, G. 1D-coupled photochemical model of neutrals, cations and anions in the atmosphere of Titan. Icarus 268, 313-339 (2016).
  \item Cavali\'e, T., Lunine, J. I., Mousis, O. A subsolar oxygen abundance or a radiative region deep in Jupiter revealed by thermochemical modeling. Nat. Astron. (2023). DOI: 10.1038/s41550-023-01928-8.
  \item Norwood, J., Moses, J., Fletcher, L. N., Orton, G., Irwin, P. G. J., Atreya, S., Rages, K., Cavali\'e, T., S\'anchez-Lavega, A., Hueso R., Chanover, N. Giant planet observations with the James Webb Space Telescope. Publ. Astron. Soc. Pac. 128, 018005 (2016).
  \item Cavali\'e, T., Hue, V., Hartogh, P., Moreno, R., Lellouch, E., Feuchtgruber, H., Jarchow, C., Cassidy, T., Fletcher, L. N., Billebaud, F., Dobrijevic, M., Rezac, L., Orton, G. S., Rengel, M., Fouchet, T., Guerlet, S. Herschel map of Saturn?s stratospheric water, delivered by the plumes of Enceladus. Astron. Astrophys. 630, A87 (2019).
  \item Rohart, F., Derozier, D., Legrand, J. Foreign gas relaxation of the J=0?1 transition of HC15N. A study of the temperature dependance by coherent transients. J. Chem. Phys. 87, 5794-5803 (1987).
  \item Dick, M. J., Drouin, B. J., Crawford, T. J., Pearson, J. C. Pressure broadening of the J=5?4 transition of carbon monoxide from 17 to 200 K: A new collisional cooling experiment. J. Quant. Spectr. Rad. Transf. 110, 619-638 (2009).
  \item Borysow, J.,Trafton, L., Frommhold, L., Birnbaum, G. Modeling of pressure-induced far-infrared absorption spectra of molecular hydrogen pairs. Astrophys. J. 296, 644-654 (1985).
  \item Borysow, A., Frommhold, L. Theoretical collision-induced rototranslational absorption spectra for the outer planets: H$_2$-CH4 pairs. Astrophys. J. 304, 849-865 (1986).
  \item Borysow, J., Frommhold, L., Birnbaum, G. Collision-induced rototranslational absorption spectra of H$_2$-He pairs at temperatures from 40 to 3000 K. Astrophys. J. 326, 509-515 (1988).
  \item Flasar, F. M., Kunde, V. G., Achterberg, R. K.,  Conrath, B. J., Simon-Miller, A. A., Nixon, C. A., Gierasch, P. J., Romani, P. N., B\'ezard, B., Irwin, P., Bjoraker, G. L., Brasunas, J. C., Jennings, D. E., Pearl, J. C., Smith, M. D., Orton, G. S., Spilker, L. J., Carlson, R., Calcutt, S. B., Read, P. L., Taylor, F. W., Parrish, P., Barucci, A., Courtin, R., Coustenis, A., Gautier, D., Lellouch, E., Marten, A., Prang\'e, R., Biraud, Y., Fouchet, T., Ferrari, C., Owen, T. C., Abbas, M. M., Samuelson, R. E., Raulin, F., Ade, P., C\'esarsky, C. J., Grossman, K. U., Coradini, A. An intense stratospheric jet on Jupiter. Nature 427, 132-135 (2004).
  \item Giles, R. S., Greathouse, T. K., Cosentino, R. G., Orton, G. S., Lacy, J. H. Vertically-resolved observations of Jupiter?s quasi-quadrennial oscillation from 2012 to 2019. Icarus 350, 113905 (2020).
  \item Twomey, S. Introduction to the mathematics of inversion in Remote Sensing and Indirect Measurements, 2nd edn. (Dower Phoenix) (2002).

\end{enumerate}

\end{document}